\documentclass[twocolumn,caps,showpacs]{revtex4}

\usepackage{graphicx,color}
\usepackage{latexsym}
\usepackage{amsmath,amssymb}        
\usepackage[draft=false]{hyperref}
\usepackage{mathrsfs}
\usepackage{comment}
\usepackage{soul}
\usepackage{mathrsfs}
\DeclareSymbolFontAlphabet{\mathrsfs}{rsfs}
\DeclareMathAlphabet{\mathcal}{OMS}{cmsy}{m}{n}

\begin{document}


\title{Parameter estimates in binary black hole collisions using neural networks}


\author{M. Carrillo}
\affiliation{Instituto de F\'{\i}sica y Matem\'{a}ticas, Universidad
              Michoacana de San Nicol\'as de Hidalgo. Edificio C-3, Cd.
              Universitaria, 58040 Morelia, Michoac\'{a}n,
              M\'{e}xico.}

\author{M. Gracia-Linares}
\affiliation{Instituto de F\'{\i}sica y Matem\'{a}ticas, Universidad
              Michoacana de San Nicol\'as de Hidalgo. Edificio C-3, Cd.
              Universitaria, 58040 Morelia, Michoac\'{a}n,
              M\'{e}xico.}

\author{J. A. Gonz\'alez}
\affiliation{Instituto de F\'{\i}sica y Matem\'{a}ticas, Universidad
              Michoacana de San Nicol\'as de Hidalgo. Edificio C-3, Cd.
              Universitaria, 58040 Morelia, Michoac\'{a}n,
              M\'{e}xico.}

\author{F. S. Guzm\'an}
\affiliation{Instituto de F\'{\i}sica y Matem\'{a}ticas, Universidad
              Michoacana de San Nicol\'as de Hidalgo. Edificio C-3, Cd.
              Universitaria, 58040 Morelia, Michoac\'{a}n,
              M\'{e}xico.}


\date{\today}


\begin{abstract}
We present an algorithm based on artificial neural networks (ANNs), that estimates the mass ratio in a binary black hole collision out of given Gravitational Wave (GW) strains. In this analysis, the ANN is trained with a sample of GW signals generated with numerical simulations. The effectiveness of the algorithm is evaluated with GWs generated also with simulations for given mass ratios unknown to the ANN. We measure the accuracy of the algorithm in the interpolation and extrapolation regimes. We present the results for noise free signals and signals contaminated with Gaussian noise, in order to foresee the dependence of the method accuracy in terms of the signal to noise ratio.
\end{abstract}




\maketitle

\section{Introduction}
\label{sec:introduction}

The recent discovery of Gravitational Waves (GWs) through the observation of the GW150914 and GW151226 events with the LIGO array \cite{LIGOMain,GW151226} opens a new window for astrophysics that will allow the study of highly dynamical processes that  have a GW component \cite{LIGOAstro}. Among the potentially observable sources of gravitational radiation, including binary neutron stars, supernovae core collapse or high energy accretion processes among others \cite{sources}, the discovery corresponds to signals due to the collision of two orbiting black holes. This is of particular importance because the astrophysical scenario is associated with a space-time without matter and in consequence, there are no physical parameters associated to high density and high temperature matter that could have been involved. In a way, this simplifies the attempt to solve the inverse problem of the cause of the GW signal, that is, the reconstruction of the parameters of the binary black hole system. 

In the end, the parameters of the source of the breakthrough cases were estimated accurately for the particular cases of GW150914 and GW151226, and reproduced with numerical relativity \cite{LIGOMain,GW151226}. It is important though to explore methods that allow the solution to the mentioned inverse problem for the general case of all the intrinsic parameters, that may eventually be more efficient. By now, due to the discovery of GWs, we consider that the most interesting astrophysical scenario is currently that of a binary black hole merger. Even though this particular problem corresponds to a vacuum space-time, the parameters characterizing the system are many and of two types, on the one hand the intrinsic parameters associated to the masses and spins of the black holes and possibly the eccentricity of the orbit; on the other hand there are extrinsic parameters related to the location of the source on the sky, including luminosity distance, declination, inclination, polarization and orbital phase of the source \cite{LALInference,LIGOParams}.

Based on these motivations, in this paper the type of GWs we analyze is sourced by the collision of two black holes and present a method to estimate intrinsic parameters. We focus on a one dimensional parameter space in order to quantify the accuracy of our method, and also as a first step toward estimating the type of samples required in more general scenarios including more intrinsic parameters. There are two interesting one parameter spaces to look at, the one of the mass ratio of the black holes and the one of the relation of aligned or anti-aligned spins of an equal mass binary system. In the general astrophysical scenario, masses and spins are both important and in principle there is no reason to choose one over the other. As specific example, if we would like to estimate the recoil velocity of the resulting kicked black hole due to the collimated emission of gravitational radiation, both, the mass ratio \cite{kick1} and non-aligned spins \cite{kick2} are important. We choose to analyze the mass ratio of the black holes, because it is a more explored case and numerically easier to implement  (see for instance the parameters in GW catalogs \cite{NINJA2,Mroue2013,PABLO2016}).

Our plan is to construct an efficient machine learning (ML) application that helps at filtering and estimating the parameters of a GW source out of a noisy signal. As a first step, what we do in this paper is to investigate whether or not an ANN can provide reliable estimates of the mass ratio of the black holes, that is, solving the inverse problem for two scenarios, one starting with noise free signals and one another starting with signals contaminated with Gaussian noise. 

On the one hand, ML methods have been applied to the GW data analysis for various purposes, for instance, in \cite{referee1} ML algorithms are used to classify and remove glitches from signals in the detectors,  whereas in \cite{referee2} they are used to classify the sensitivity improvements in terms of the masses of the BBHs. On the other hand, the parameter estimate of a GW source counts with efficient inverse problem strategies, including for instance the use of Markov chains \cite{Montecarlo} and interpolation of waveforms \cite{Interpolation}, which have shown to be very efficient. In the LALInference library, the MCMC methods use stochastic wandering through the parameter space \cite{LALInference}. Even though we have not attempted to apply our method to a bigger parameter space yet, we expect the ANNs, using a selected set of points to be more efficient at classifying and estimating parameters than wandering, because the efficiency becomes more relevant when the number of parameters increases.

Using ANNs to track down a parameter needs three steps. First, training the ANNs uses a set of GW signals associated to a given mass ratio, which will fix the topology and weights of the network. Second, the network is validated using a different set of signals. The validation determines the number of iterations required to provide accurate predictions and avoid overtraining the network. The third step is the prediction of the mass ratio given the GW strain unknown to the ANN.

The sample of GW signals used to train, validate and measure the accuracy in prediction of the ANN, is generated using numerical simulations of black hole collisions. For this we use the Einstein Toolkit (ETK) code \cite{ETK,ETK2}. There are catalogs with collections of GW strains, constructed either with numerical relativity simulations (e.g.\cite{NINJA2,Mroue2013,PABLO2016}), with phenomenological models \cite{SASCHA} or surrogate models \cite{surrogate}, that include GW strains containing orbital, plunge and ringdown phases. Even though, we decided to generate our own signals because we can define the size and distribution of our sample. In exchange of having control of the sample, we sacrifice the orbital phase of the strain; nonetheless the relevance of our result relies on the presentation of the method that can be replicated with any set of GWs.

The paper is organized as follows. In section \ref{sec:description} we present an overview of the method that we employ to estimate the mass ratio of the black holes. In section \ref{sec:waveforms} we describe how to prepare the waveforms. Section \ref{sec:nn} is a general description of artificial neural networks. In section \ref{sec:results} we present the results for noise free signals and signals contaminated with Gaussian noise. Finally, in section \ref{sec:final} we draw some conclusions and perspectives.

\section{Description of the method}
\label{sec:description}

The algorithm to estimate the mass ratio $q$ from the waveforms is outlined below:

\begin{itemize}

\item{Generation of the waveforms: Solve numerically Einstein's equations for 32 different initial data representing values of the mass ratio in the range between $1$ and $5$. The result of this step is a set of 32 files containing the real part of the $l=2,~m=2$ mode of the gravitational wave strain, namely $h_{+}$ and the corresponding mass ratio resulting from each simulation (for details see section \ref{sec:waveforms}). }

\item{Selection of the sets: Divide the set of 32 GWs in three smaller sets. Training set (10 GWs), Validation set (3 GWs) and Prediction set (19 GWs) as presented in Table \ref{tab:sets}. The first set is used to train the ANNs, the second one is used to avoid overfitting and decide when to stop the training and the third one allows us to test the prediction accuracy of the implementation.
The main difference between the training and prediction sets is that the former contains GWs with mass ratio in the range from $1$ to $2.5$ and the later contains GWs from the whole interval from 1 to 5. This allows us to study the interpolation and extrapolation capabilities of the algorithm.}

\item{Structure of the ANN: Construct the network with 3 layers. The number of neurons in the input layer is 130, in the hidden layer between 10 and 100 and in the output layer only one neuron. The input neurons will receive the 130 first values of the rescaled strain starting from its maximum. The output of the network will be compared with the desired value of the mass ratio associated with the corresponding strain. All the neurons are connected by weights initialized randomly. These values have to be updated using a training method to obtain the desired outputs. Details in section \ref{sec:nn}}

\item{Training and validation of the ANN: Present every element of the training set to the network and minimize the difference between the desired and the obtained output changing the values of the weights. At the same time, present the elements of the validation set to the network without minimizing the error, i.e. only validating the accuracy of the algorithm. Continue with this process until a given tolerance is obtained or when the errors using the validation set start growing. When the training is finished, the corresponding weights represent the network that will be used for prediction.}

\item{Prediction with the ANN: Finally, use the best weights of the ANN from the training process to evaluate the prediction set and establish the accuracy of the method to predict values in the interpolation and extrapolation regimes.}

\end{itemize}

\section{Generation of the waveforms}
\label{sec:waveforms}

As mentioned before, among the intrinsic parameters describing a black hole collision, we consider only the mass ratio between the black holes. We produce GW strains using numerical simulations generated with the ETK \cite{ETK,ETK2}, considering the initial separation between black holes to be that of the first quasi-circular orbit \cite{qc0} in order to obtain a quick merger for various mass ratios between the two black holes, ranging from 1 to 5. For this we use the BSSN formulation, a moving puncture gauge  and moving boxes that track the location of the punctures before and after merger \cite{movingpunctures, movingpunctures2}. The numerical domain uses seven refinement levels with the Carpet finite differences driver \cite{carpet}. In our simulations, we use the domain $[-120M,120M]^3$ with bitant symmetry. We have verified that in all the initial configurations the ADM mass of the space-time is the same. 

\begin{figure}
\centering
\includegraphics[width= 8cm]{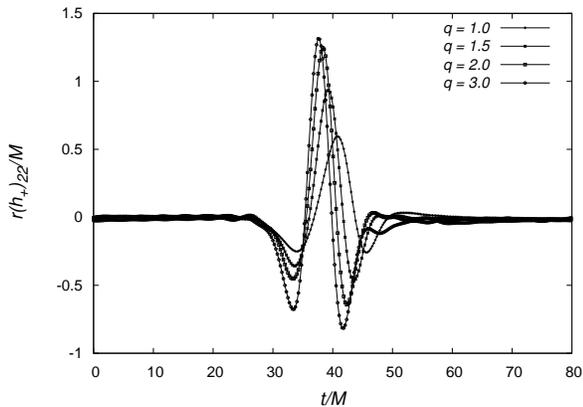}
\caption{We present the real part of the $l=2,~m=2$ mode of the strain $h_{+}$ for four representative cases. These signals are appropriately rescaled with the extraction radius ($r$) and the ADM mass of the system ($M$).}
\label{fig:gw_training}
\end{figure}

At radius $r=30M$ we extract the $l=2$, $m=2$ mode of the $\Psi_4$ Weyl scalar. We post-process this measure to convert it to the real part of the strain $(h_{+})_{22}$ following the recipe in \cite{ReisswigPollnew2011}. For simplicity we will use indistinctly $(h_{+})_{22}$ or $h_{+}$ as our signal.  In Fig. \ref{fig:gw_training} we show the real part of the strain for a subset of cases with different mass ratios. As mentioned before, we decide to choose the input to the ANN to be a set of values of $h_{+}$ in a given time window. For this we set the output time resolution to $0.125M$ and the data size will be described in detail below.

\begin{table}
\begin{tabular}{|c|c|c|}\hline
Training & Validation	& Prediction \\\hline\hline
1.00 & 2.75 & 1.2, 1.4 \\
1.10	& 3.00	& 1.6, 1.8 \\
1.25	& 3.25	& 1.9, 2.1\\
1.30	& 		& 2.2, 2.3\\
1.50	&		& 2.4, 2.6\\
1.70	&		& 2.7, 2.8\\
1.75	&		& 2.9, 3.5\\
2.00	&		& 3.4, 3.75\\
2.25	&		& 4.0, 4.5\\
2.5	& 		& 5.0\\\hline
\end{tabular}
\caption{\label{tab:sets} Training, validation and prediction sets used for the three different stages of the ANN method. The numbers correspond to the mass ratio $q$ of the black holes.}
\end{table}

\section{Neural network description}
\label{sec:nn}

ANNs arose as an attempt to emulate the nervous system in both organization and stimuli processing from internal and external environment of living organisms \cite{Russell}. Basically, a neural network is fed with some variables or parameters of a particular problem, and after a series of calculations it delivers a concrete outcome correlated with the input data. This kind of artificial intelligence is commonly used in pattern classification and recognition, which makes it specially appropriate to analyze the problem we intend to engage here.

\subsection{Mathematical Representation of an ANN}

\begin{figure}
\centering
\includegraphics[width = 8.5cm]{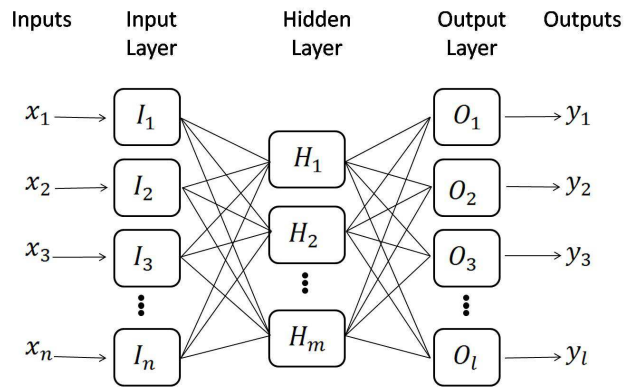}
\caption{Structure of a 3 layer feedforward ANN. The links between the $I_i$ and the $H_j$ are weighted by the coefficient $w_{ij}$ and the weights between $H_j$ and $O_k$ are $\tilde{w}_{jk}$. }
\label{fig:my_label}
\end{figure}

In order to illustrate the operation of a neural network, consider a 3 layer feedforward network with $n$ inputs, $m$ hidden neurons and $l$ outputs as illustrated in Fig. \ref{fig:my_label}, where we have $n * m + m * l$ weights. Since in the input layer no calculations are made, the $j$-th hidden neuron has the input vector $\{x_{1},\dots,x_{i},\dots,x_{n}\} $ and its corresponding weight vector $\{w_{1j},\dots,w_{ij},\dots,w_{nj}\}$, where $i$ labels  an input neuron. Then the transfer function maps the input vector from $R^n$ to $R$, taking the linear combination of input vectors with the elements $w_{ij}$ as coefficients:

\begin{equation}
\eta_j = \sum_{i=1}^{n} w_{ij}x_{i}  + w_{j0} \, ,
\end{equation}
where $w_{j0}$ is the weight of an extra synapse called \textit{bias} with constant input equal to one, added to shift the transfer function result or to avoid the particular case of a zero vector. Immediately after, the activation function $F$ enters in action giving the output $\sigma_j$:
\begin{equation}
 \sigma_j = F(\eta_j) = F\left(\sum_{i=1}^{n} w_{ij}x_{i} + w_{j0}\right).
\end{equation}

Similarly for the $k$ output neuron we have the result:

\begin{equation}
y_k =  G(\zeta_k) = G \left(\sum_{j=1}^{m} \tilde{w}_{jk}\sigma_j  + \tilde{w}_{k0} \right) \, ,
\end{equation}

\noindent where $G$ is the activation function for the output layer neurons, $\tilde{w}_{jk}$ are the weight coefficients corresponding to the input vector of the $k$ output neuron, $\tilde{w}_{k0}$ its bias and $\zeta_k$ is the linear combination of $\sigma_j$ with coefficients $\tilde{w}_{jk}$.

\subsection{Selection of the inputs and topology of the network}

As mentioned before, we include the training, validation and prediction phases. In Table \ref{tab:sets} we indicate the sets chosen for each of them. The extraction of the physical properties of the GW strain relies on an adecuate selection of the inputs that will be introduced to the ANN. Given a signal $h_{+}$, we identify its maximum amplitude and starting from this point we select the next 130 values of the signal. We use these 130 values as inputs for our neural networks. This part of the signal contains information of the merger and ringdown phases. In fact, choosing the maximum as a reference point to define a time window for the analysis has been useful in previous analyses \cite{LALInference}. The number 130 was selected after testing different time window sizes, and finding this one contained enough information as to provide accurate results.

The $h_{+}$ signals are rescaled to the range $(-1,1)$, using as reference the maximum and minimum of the strain within the training set. As the rule of thumb suggests, the neural network is initialized on each neuron with random weight values between $(-1/ \sqrt{d}, 1/ \sqrt{d} )$, with $d$ the number of income connections for that neuron. Each hidden neuron is set to an hyperbolic tangent activation function instead of the standard logistic function \cite{LeCun}, whereas the output neuron is set with a linear function restricted only to positive values. This selection was found to provide the smallest error during the training phase.
	
About the structure of the neural network. Since there is no rule about the number of hidden neurons, we perform the process with different numbers of hidden neurons ranging from 10 to 100. The number of output neurons is one, corresponding to the unique parameter we want to predict, the mass ratio.

Another important ingredient for the prediction is how the network is initialized. In our approach, we do not simply use a single ANN but use a set of 10 networks with the same topology and different initial weights chosen randomly and in the end calculating an average of the prediction among them.

\subsection{Training}

In order to obtain an accurate value as an  expected output, one must get the appropriate configuration and the ANN weights. The procedure to do this consists in changing all the synaptic weights during the  training phase. Suppose we have a set of $N$ pairs consisting of a signal $h_{+}$ and the associated mass ratio $q$, that is, a vector ${h}^{p}_{+}$ with 130 entries, and a scalar $\tilde{q}^p$, $\{ {h}^{p}_{+}, \tilde{q}^p \}$, where $p$ labels a given pair. During the training phase, the objective is to reduce the error between the result $q^p$ predicted by the ANN and the truly desired output $\tilde{q}^p$.

The most common way to measure this difference is by adding the squared errors from output neurons:

\begin{equation}
E^p = \frac{1}{2} \sum_{k} {\left(\tilde{q}_{k}^p - {q}_{k}^{p} \right)}^2 \, ,
\end{equation}

\noindent where $k$ is the index of the output neurons, in our case is only one. The overall performance or global error is defined by:

\begin{equation}
E^G = \sum_p E^p = \sum_p  \sum_{k} \frac{1}{2} {\left(\tilde{q}_{k}^p - q_{k}^{p} \right)}^2 \, .
\end{equation}

The method we use to minimize this error is back-propagation \cite{backpropagation}. In the implementation used in this paper, the training took between ten and seventy thousand iterations when the number of hidden neurons was selected between 10 to 100. Each iteration consists in presenting the ten elements of the training set in Table I to the network and updating the weights to minimize the error. The time this process takes depends on the number of hidden neurons and ranges from 3 minutes using 10 neurons to 1 hour using 100 neurons using one processor. The training phase is stopped when the validation error reaches the global minimum  as shown in Figure \ref{fig:training_validation_error}.
To calibrate the accuracy of the network during training, we estimate the number of correct classified GWs within a given error. Using 10 hidden neurons the accuracy of the network after training, evaluated with the whole training set is 100\% within 2\% error, whereas using 100 neurons the error is within 0.5\%.

\subsection{Validation}

\begin{figure}
\centering
\includegraphics[width=0.5\textwidth]{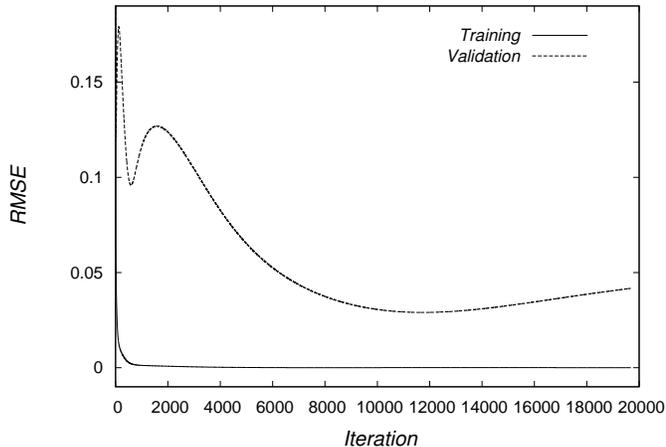}
\caption{Training and validation error as a function of the number of iterations for a network with 10 hidden neurons. As mentioned in the text, when the global minimum of the error in the validation set is reached (around 11000 iterations in this example) the training is stopped.}
\label{fig:training_validation_error}
\end{figure}

We use the validation set in Table \ref{tab:sets} to verify that the weights calculated during training provide accurate predictions for an unknown set of waveforms to the network. For this we take the weights of a network for each iteration during training, and use those weights to classify the inputs from the validation set. In Fig. \ref{fig:training_validation_error} we show the normalized root mean square error (RMSE) for the training set compared to the RMSE for the validation set as a function of the number of iterations. The RMSE is defined by:

\begin{equation}
RMSE = \sqrt{     \sum^{N}_{p=1}{ \frac{\left(\tilde{q}^{p} - {q}^{p} \right)^2}{N}} }\, ,
\label{eq:MSE}
\end{equation}

\noindent where $\tilde{q}^{p}$ and $q^p$ are the real and the estimated value of the mass ratio with index $p$, which labels each of the 10 training cases or each of the 3 validation cases in Table \ref{tab:sets}.

\subsection{Prediction}

Finally, the prediction set is split into two regimes. The interpolation regime includes the mass ratios $q$ running from 1.2 to 2.4, and the extrapolation regime includes the values of $q$ from 2.6 until 5.0. As shown below, the prediction accuracy in these two regimes is completely different.

\section{Results}
\label{sec:results}

\subsection{Noise free strain}

In Table \ref{tab:TableOutcomes10HN} we present the predicted mass ratios using ANNs with 10, 50 and 100 hidden neurons. In the Table, the average in the prediction is calculated among the 10 networks, remember that for each case we set ten networks in order to smooth out random initialization problems.

\begin{figure}
\centering
\includegraphics[width=0.5\textwidth]{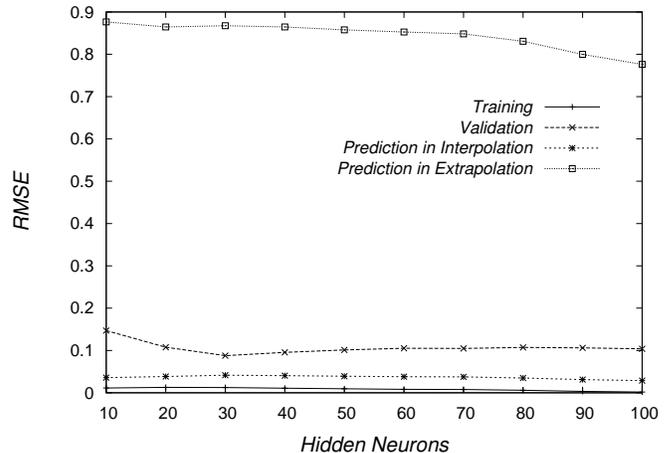}
\caption{Training, validation and prediction error as a function of the number of hidden neurons. Training error is always smaller than the validation error as expected. The prediction error is smaller in the interpolation set than in the extrapolation regime. The errors in all the cases are pretty independent of the number of hidden neurons.}
\label{fig:TrainingValidationError}
\end{figure}

In Figure \ref{fig:TrainingValidationError} we show the training, validation and prediction RMSE as a function of the number of hidden neurons. The error in the prediction set has been divided into the interpolation and extrapolation regimes. Notice that each of these errors is nearly independent of the number of hidden neurons. This is important information that will be useful in more complicated physical scenarios involving higher dimensional parameter spaces.

\begin{figure}
	\centering
	\includegraphics[width=0.5\textwidth]{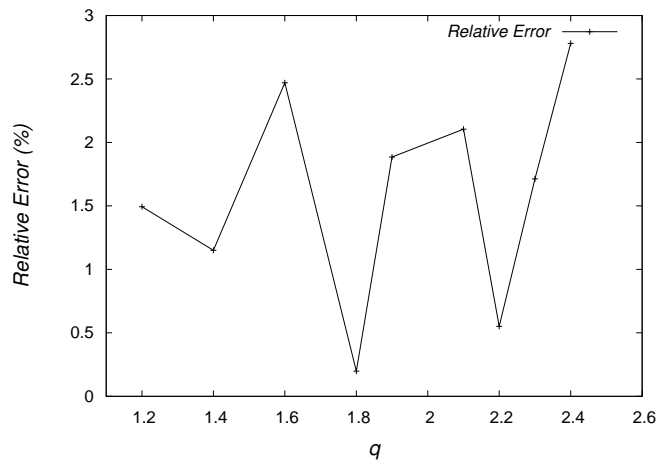}
	\includegraphics[width=0.5\textwidth]{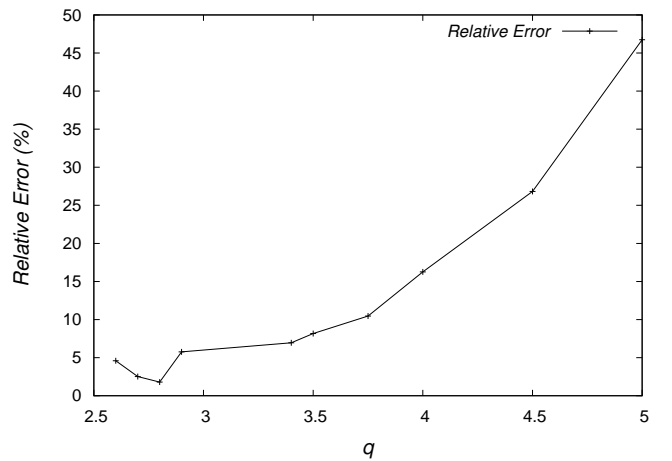}
	\caption{Percent of uncertainty as a function of the mass ratio estimated with ANNs using 10 hidden neurons in the interpolation (top) and the extrapolation (bottom) regimes.}
	\label{fig:RelativeErrorInterpolation}
\end{figure}

\begin{figure}
\centering
\includegraphics[width=8.5cm]{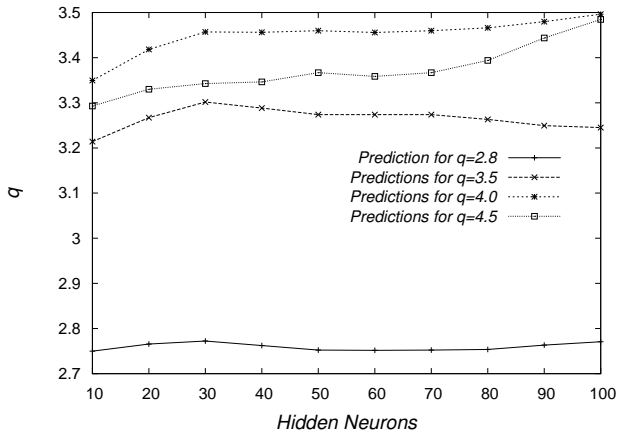}
\caption{Predicted values of the mass ratio for various extrapolation cases, as a function of the number of hidden neurons. The accuracy is not as good as in the interpolation case, nonetheless the predicted value does not seem to depend essentially on the number of hidden neurons. By increasing this number, perhaps it may happen that the prediction improves as seems to happen for $q=4.5$ at the price of increasing the computer resources.}
\label{fig:MassRatio}
\end{figure}

In Fig. \ref{fig:RelativeErrorInterpolation} we present the error of the predicted mass ratios using an ANN with 10 hidden neurons, in the interpolation and extrapolation regimes. In the interpolation regime the error remains small and very stable whereas in the extrapolation regime the error clearly grows as the mass ratio departs more and more from the training set values. 

In order to explore whether or not increasing the number of neurons helps at decreasing the error in the prediction regime, we present in Fig. \ref{fig:MassRatio} the predicted value as a function of the number of hidden neurons, showing that in all the cases of our sample the improvement is not significant.

\begin{table}
\centering
\begin{tabular}{ |c|c|c|c| }
\hline
\multicolumn{4}{|c|}{Predicted $q$ by the ANN} \\ \hline
Target $q$ & ANN-10 & ANN-50 & ANN-100\\
\hline
{1.2} &  
$1.182 \pm 0.003$ &  
$1.183	\pm 0.0001$ &
$1.186 \pm 0.001$ \\ 
\hline
{1.4} &
$1.384 \pm	0.007$ &  
$1.384 \pm 0.0008$ &
$1.399 \pm 0.004$\\
\hline
{1.6} &
$1.560 \pm 0.008$ &  
$1.593 \pm 0.003$ &
$1.621 \pm	0.007$\\
\hline
{1.8} &
$1.796 \pm 0.003$ &  
$1.799 \pm 0.001$ &
$1.798 \pm 0.002$\\
\hline
{1.9} &
$1.936 \pm 0.005 $ &  
$1.930	\pm 0.001$ &
$1.920 \pm 0.004$\\
\hline
{2.1} &
$2.056 \pm 0.009$ &  
$2.042 \pm	0.002$ &
$2.059 \pm 0.005$\\
\hline
{2.2} &
$2.188 \pm 0.001 $ &  
$2.192 \pm 0.0007$ &
$2.193 \pm 0.0004$\\
\hline
{2.3} & 
$2.261 \pm 0.004$ &  
$2.263 \pm 0.001$ &
$2.270 \pm 0.002$\\
\hline
{2.4} &
$2.333 \pm 0.012$ &  
$2.314 \pm 0.003$ &
$2.338 \pm 0.007$\\
\hline
{2.6} &
$2.481 \pm 0.013$ &  
$2.475 \pm 0.003$ &
$2.514 \pm 0.013$\\
\hline
{2.7} &
$2.632 \pm 0.012 $ &  
$2.639 \pm 0.001$ &
$2.667 \pm 0.012$\\
\hline
{2.8} &
$2.750 \pm 0.016 $ &  
$2.756 \pm 0.004$ &
$2.771 \pm 0.008$\\
\hline
{2.9} &
$2.733 \pm 0.023 $ &  
$2.788 \pm 0.003$ &
$2.825 \pm 0.013$\\
\hline
{3.4} & 
$3.163 \pm 0.032$ &
$3.229 \pm 0.008$& 
$3.216 \pm 0.004$\\
\hline
{3.5} &
$3.214 \pm 0.036 $ &  
$3.278 \pm 0.013$ &
$3.245 \pm 0.013$\\
\hline
{3.75} &
$3.358 \pm 0.035$ &
$3.441 \pm 0.012$ & 
$3.432	\pm 0.009$ \\
\hline
{4.0} &
$3.350 \pm 0.034$ &  
$3.455 \pm 0.010$ &
$3.496 \pm 0.007$\\
\hline
{4.5} &
$3.293 \pm 0.029$ &  
$3.353 \pm 0.007$ &
$3.485 \pm 0.032$\\
\hline
{5.0} &
$2.663 \pm 0.090$ &  
$2.647 \pm 0.022$ &
$2.873 \pm 0.063$\\
\hline
\end{tabular}
\caption{Averaged predictions calculated with ten randomly initialized neural networks in each case, using 10, 50 and 100 hidden neurons for the signals corresponding to the prediction set.}
\label{tab:TableOutcomes10HN}
\end{table}

\subsection{Strain with Gaussian Noise}

\begin{figure}
	\centering
	\includegraphics[width=8.5cm]{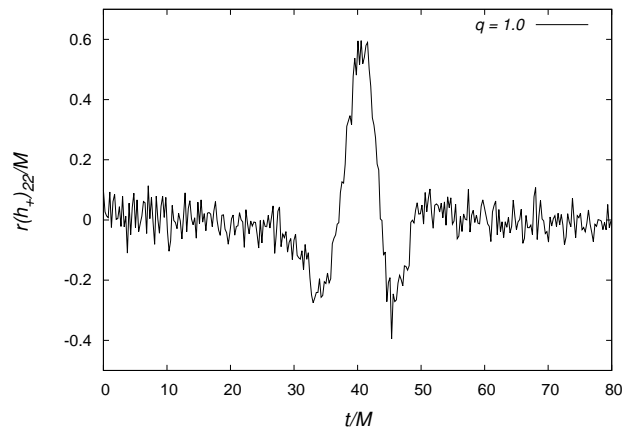}
	\caption{Scaled $h_{+}$ for the equal mass case, contaminated with Gaussian noise such that the SNR is 10.}
	\label{fig:noisysignal}
\end{figure}

We now study the accuracy of the method using noisy strains. As a first step we consider the use of Gaussian noise before tackling more complex types of noise. We add this noise to the signals $h_{+}$ of the previous subsection with noise amplitudes corresponding to signal to noise ratios from 0.5 to 25. This contamination was added to each of the strain signals of the training, validation and prediction sets. In Fig. \ref{fig:noisysignal} we present the strain for the case $q=1$ and SNR=10.

\begin{figure}
	\centering
	\includegraphics[width=8.5cm]{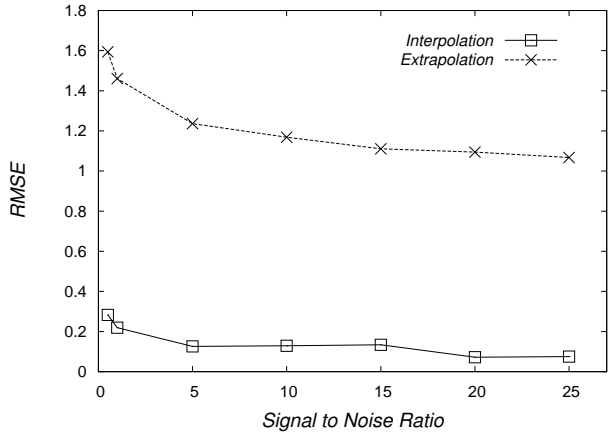}
	\caption{Root mean square error for the prediction set using the 10 ANNs with 10 hidden neurons, using SNR in the range from 0.5-25, for both the interpolation and extrapolation regimes.}
	\label{fig:SNR_RMSE}
\end{figure}

In Fig. \ref{fig:SNR_RMSE} we show the RMSE for the interpolation and extrapolation regimes. As expected, the error decreases as the SNR increases. Notice that  the magnitude of the errors is bigger when the analysis is done with noisy signals as can be seen by comparing these errors with those in Fig. \ref{fig:TrainingValidationError}. It is clear that the accuracy of the network decreases as we increase $q$ (as in the previous subsection) but also decreases when we decrease the SNR. This behaviour is expected because as the SNR decreases the ANN gets a hard time trying to find the signal inside the noise. One way we believe this can be improved is using a convolution of the noisy strain as a filter to increase the size of the signal compared with the noise.

\section{Final comments}
\label{sec:final}

We explored the accuracy in solving the inverse problem for the estimate of the mass ratio with ANN methods, for which we used a family of numerically generated GW signals from the collision of two black holes. We estimate the accuracy in the prediction in terms of the parameters of the ANN such as the number of hidden neurons. It is also expected that the accuracy of the method improves when the time window of the signal contains information of the orbiting phase.

Our results indicate that the training set  provides accurate predictions within $3\%$ error for the interpolation regime. For values of $q$ in the extrapolation regime the error growths very fast. The conclusion we can obtain from this, is the need of expanding the training sets to avoid as much as possible, the extrapolation regime. We have shown that our method is accurate even with a modest sample of signals in the interpolation regime. Been accurate with small samples is particularly interesting now that the parameter space including aligned, anti-aligned and precessing black hole spins accounts with more GW signals \cite{PABLO2016}, where our method -with this accuracy- could be useful to classify both spins and mass ratios.

In terms of the results related to the analysis with noisy signals, the results are promising. Nevertheless the method can improve in order to obtain better results, which may include pre-processing of the signals using specific filters.


\section*{Acknowledgments}

This research is supported by grants CIC-UMSNH-4.9, CIC-UMSNH-4.23  and CONACyT 258726 (Fondo Sectorial de Investigaci\'on para la Educaci\'on).


\end{document}